# CoRoT-2a magnetic activity: hints for possible star-planet interaction

Isabella Pagano[1], Antonino F. Lanza[1], Giuseppe Leto[1], Sergio Messina[1], Pierre Barge[2], Annie Baglin[3]

*[1] INAF, Osservatorio Astrofisico di Catania, Italy; [2] Laboratoire d'Astrophysique de Marseille, France; [3] LESIA, Observatoire de Paris, France*

+39-09-7332243

+39-095-2937823

isabella.pagano@oact.inaf.it

CoRoT-2a is a young (≈0.5 Gyr) G7V star accompanied by a transiting hot-Jupiter, discovered by the CoRoT satellite (Alonso et al. 2008; Bouchy et al. 2008). An analysis of its photospheric activity, based on spot modelling techniques previously developed by our group for the analysis of the Sun as a star, shows that the active regions on CoRoT-2a arised within two active longitudes separated by about 180 degrees and rotating with periods of 4.5221 and 4.5543 days, respectively, at epoch of CoRoT observations (112 continous days centered at ≈2007.6). We show that the total spotted area oscillates with a period of about 28.9 days, a value close to 10 times the synodic period of the planet with respect to the active longitude pattern rotating in 4.5221 days. Moreover, the variance of the stellar flux is modulated in phase with the planet orbital period. This suggests a possible star-planet magnetic interaction, a phenomenon already seen in other extrasolar planetary systems hosting hot-Jupiters.

*planetary systems; hot-Jupiters*

SPI: Star Planet Interaction

## Introduction

Among the 342 extrasolar planets discovered up to March 2009, about 25% are massive planets ($M_p \sin i > 0.2\ M_J$) in tight orbits (< 0.1 AU) around their parent stars. Cuntz et al. (2000) predicted that a giant planet in a close orbit may increase stellar activity by means of tidal and magnetospheric interactions. Observational support to such a Stellar-Planet Interaction (SPI) has been provided by several authors: e.g., Shkolnik et al. (2003, 2005, 2008) found chromospheric emission in several stars hosting planets to be variable with the planet orbital period; Henry et al. (2002) and Walker et al. (2008) found hints for photospheric activity modulation in phase with the planet orbital period.

Several other independent studies have highlighted that a short-period planet can induce activity in the photosphere and upper atmosphere of its host star. The nature of SPI appears to be strongly affected by both the stellar and planetary magnetic fields. Shkolnik et al. (2009) and Schmitt (2009) show that for extrasolar planets around young stars conditions similar to those observed for the Jupiter-Io system are met and consequently similar processes are expected to occur on far larger scales. Lanza (2008) describes the SPI considering the reconnection between the stellar coronal field and the magnetic field of the planet. Reconnection events produce energetic particles that, moving along magnetic field lines, impact onto the stellar chromosphere giving rise to a localized hot spot. Moreover, reconnection events in the corona might affect subphotospheric dynamo action in those stars producing localized photospheric (and chromospheric) activity migrating in phase with their planets.

Long baseline and continuous observations provided by CoRoT (Baglin et al. 2006) are optimal to study the photospheric variability of the stars hosting the planets it detects. CoRoT has recently discovered a hot-Jupiter, CoRoT-2b, orbiting around a main-sequence G7 star which displays a remarkable photospheric activity (Alonso et al. 2008; Bouchy et al. 2008). A detailed spot modelling analysis of the light curve of CoRoT-2a has been performed by Lanza et al. (2009). Here we focus on the aspects suggesting the presence of a star-planet magnetic interaction.

## Observations

CoRoT-2a was observed from May 16 to October 5, 2007. We extracted from the data archive the N2 chromatic light curves (Baudin et al. 2006) having a sampling of 512 s during the first week and 32 s thereinafter. The white colour light curve shown in Fig. 1 is obtained by the combination of red, green and blue fluxes (see Lanza et al 2009 for further details on data reduction). Transits and rotational modulation due to stellar activity are clearly visible.

## Results

### Active regions derived by spot modelling

The high-quality data allow us to model the evolution of the active regions with a resolution time of 3.15 days. In Fig. 2 we show the normalized spot filling factor versus longitude and time, obtained by a maximum entropy spot modelling method (see Lanza et al. 2009 for details). The star shows two active longitudes, one fixed in the adopted reference frame rotating with a period of 4.5221 days, the other slowly migrating, which corresponds to a rotation period of 4.5543 days. Signature of the two active regions can be easily found also in the light curve folded with the stellar rotation period, i.e. 4.5221 days, when flux is averaged in phase bins having a width of 0.05, as shown in Fig. 3.

### Star-Planet Interaction

In order to search for light curve variability possibly linked to the planet orbit, we have folded the times series with the planet ephemeris - $P_p$=1.7429964 d, HJD0=2454237.5356 after Alonso et al. 2008 - and computed the average flux and its variance in phase bins of 0.05. A clear wave-like behavior is apparent in the flux variance vs. orbital phases as shown in Fig. 4. The maximum variability is observed when the planet is in front of the star, slightly before the planet transit, and the minimum when it is behind. We have analised the flux variance for the first 75 days of observations only (i.e., about 43 planet orbits), because the subsequent subset is useless for our analysis being strongly affected by jitter instrumental effects that produce outliers (easy to remove) and an increased noise that pollutes the flux variance.

A further suggestion for SPI comes from the analysis of the total spotted area, plotted vs. time in Fig. 5 which shows a cyclic oscillation with a period of 28.9 ± 4.8 (1σ) days. Such a period is close to 10 times the synodic period of the planet with respect to the active longitude pattern rotating in 4.5221 days ($P_{syn}$= 2.836d).

## Discussion

The suggested SPI can be explained by the model proposed by Lanza (2008). Specifically, if some subphotospheric dynamo action takes place in the star, the amplified field emerges once the radial gradient of its intensity reaches some threshold value. The emergence of the flux might be triggered by a small localized perturbation of the dynamo effect associated with the planet when it passes by the most active longitude. An integer number of synodic periods are necessary for the planet to trigger again field emergence in a given active longitude because the planetary perturbation can play a role only when the field is already close to the threshold thanks to the steady amplification provided by the stellar dynamo.

**Acknowledgments:** The present study is based on observations obtained with CoRoT, a space project operated by the French Space Agency, CNES, with participation of the Science Program of ESA, ESTEC/RSSD, Austria, Belgium, Brazil, Germany and Spain. AFL, IP, GL and SM have been partially supported by the Italian Space Agency (ASI) under contract ASI/INAF I/015/07/0, WPP3170.

**References**

Alonso, R., et al 2008, A&A 482, L21

Baglin, A., Auvergne, M., Boisnard, L., et al. 2006, 36th COSPAR Scientific Assembly, 36, 3749

Baudin, F., et al., 2006, ESA SP-1306, Editors: M. Fridlund, A. Baglin, J. Lochard and L. Conroy. ISBN 92-9092-465-9, p.145

Bouchy, F. et al. 2008, A&A 482, L25

Cuntz, M. et al. 2000, ApJ 533, L15

Henry, G.W. et al. 2002, ApJ *577, L111*

Lanza, A.F. 2008, A&A 487, 1163

Lanza, A. F., Rodonò, M., Pagano, I., Barge, P., Llebaria, A. 2003, A&A 403, 1135

Lanza, A. F., Pagano, I., Leto, G. et al., 2009, A&A, 493, 193

Schmitt, J.H.M.M., AIP Conf. Proc, Vol 1094, pp. 473, DOI:10.1063/1.3099151

Shkolnik, E. et al. 2003, ApJ *597,1092*;

Shkolnik, E. et al. 2005, ApJ *622, 1075*

Shkolnik, E. et al. 2008, ApJ *676, 628*

Shkolnik, E.,; et al. 2009, AIP Conf. Proc, Vol 1094, 275, DOI:10.1063/1.3099102

Walker, G.A.H. et al. 2008, A&A *482, 691*

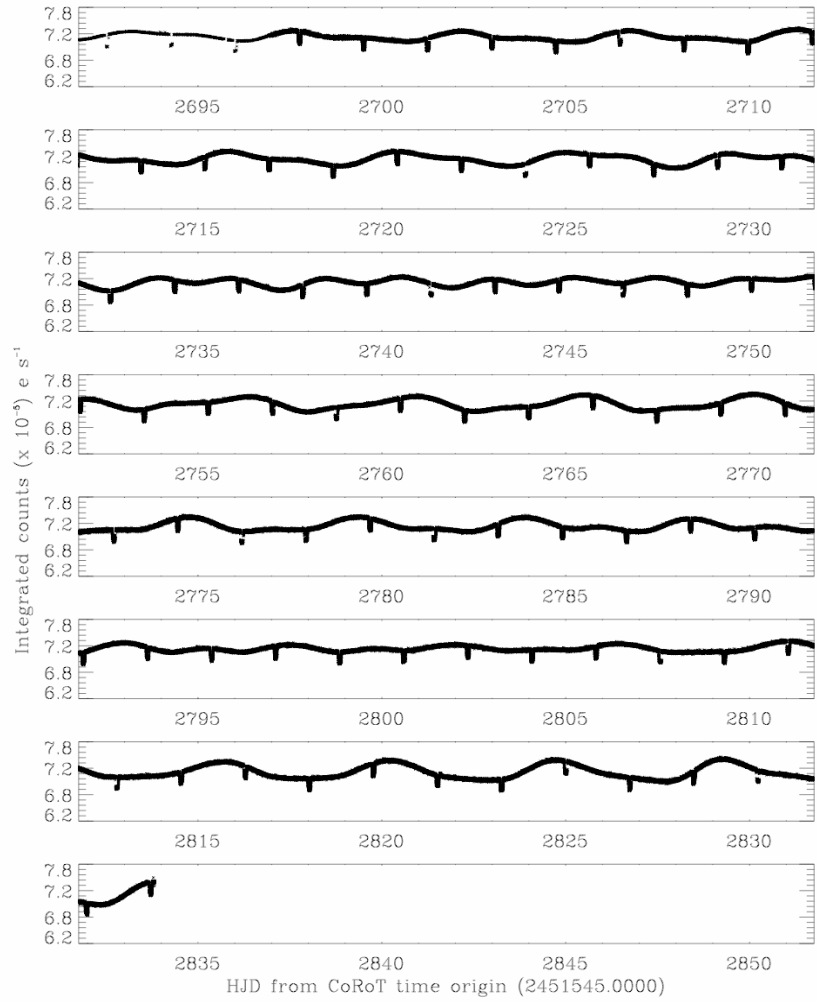

Figure 1. CoRoT-2ab white light time series. Time is measured in HJD since 2,450,000.

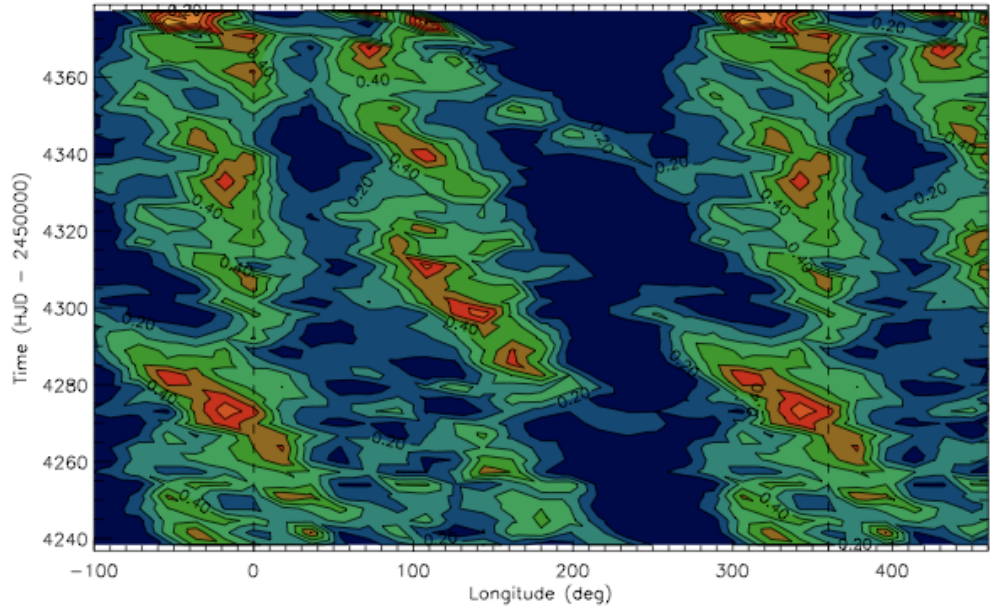

**Figure 2. Isocontours of the ratio $f/f_{max}$ where $f$ is the spot covering factor and $f_{max}$ its maximum value, vs time and longitude (from Lanza et al. 2009). Time is measured in HJD since 2,450,000.**

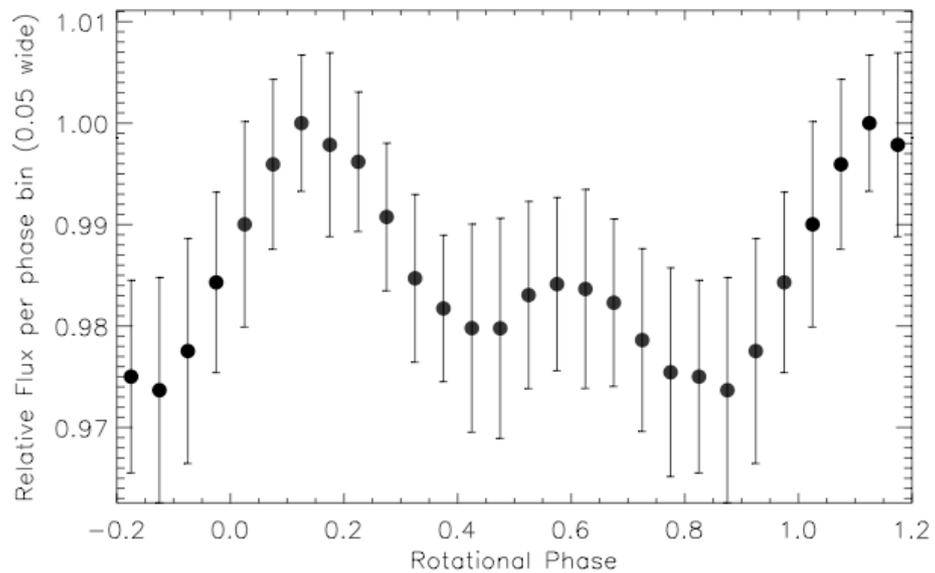

**Figure 3. Flux folded with the stellar rotation period (P=4.5221 days). The minima at phases ≈0.47 and ≈0.85 are due to the two main spotted area whose evolution can be followed in Fig.1.**

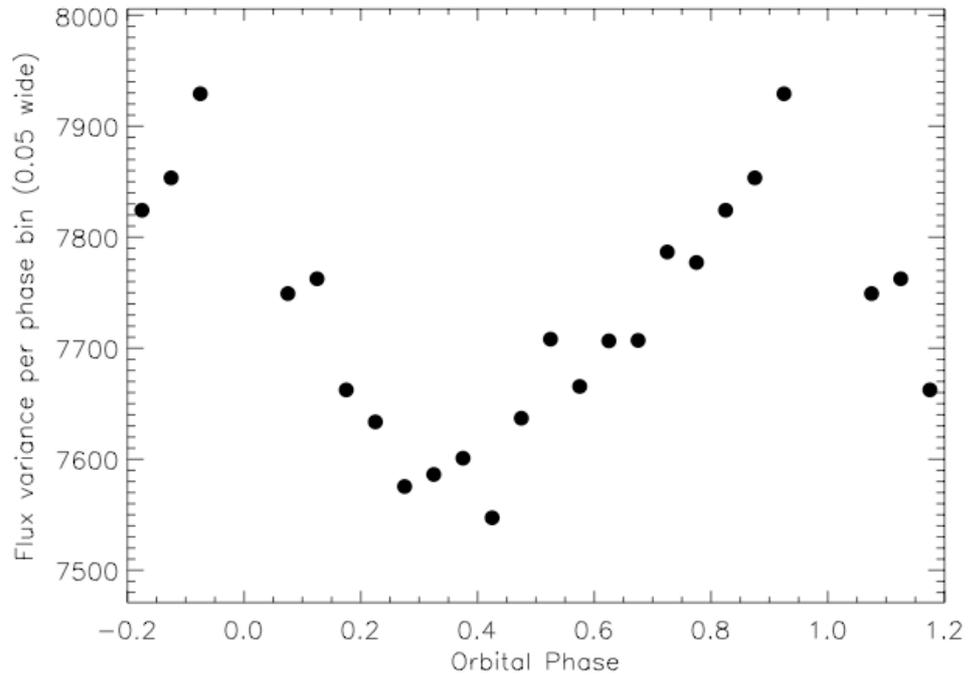

Figure 4. Flux variance vs planetary orbital phase (in bins of width 0.05). Phases are reckoned by means of the planet ephemeris P=1.7429964 d, HJD0=2454237.5356 after Alonso et al. 2008.

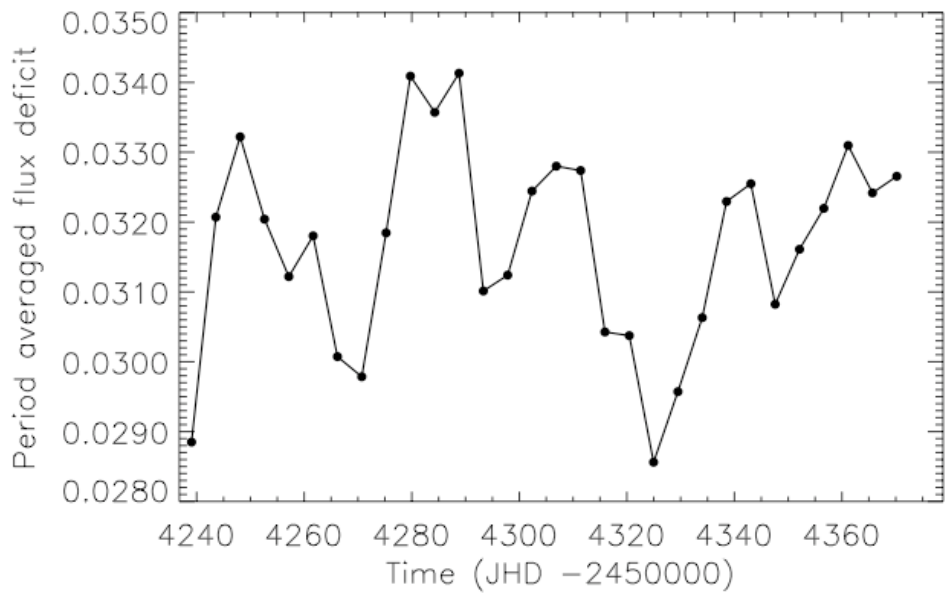

Figure 5. Variation of the total spotted area vs time from the ME spot models. The area is measured in units of the stellar photosphere. The error bars account only for random errors in the area and have a semi-amplitude of 3 standard deviations.